\documentclass[a4paper,11pt]{article}
\pdfoutput=1 

\usepackage{jheppub} 

\usepackage[T1]{fontenc} 
\usepackage{tikz}
\usepackage{bm}       
\usepackage{amssymb}  
\usepackage{amsmath}
\usepackage[utf8]{inputenc}
\usepackage{mathrsfs}
\usepackage{hyperref}
\usepackage{color}
\usepackage[normalem]{ulem} 
\usepackage{braket}

\title{\boldmath Kinks and realistic impurity models in $\varphi^4$-theory}

\author[a,b]{Mariya A. Lizunova,}
\author[b]{Jasper Kager,}
\author[b]{Stan de Lange,}
\author[b]{Jasper van Wezel}

\affiliation[a]{Institute for Theoretical Physics, Utrecht University, Princetonplein 5, 3584 CC Utrecht, The Netherlands}
\affiliation[b]{Institute for Theoretical Physics Amsterdam, University of Amsterdam, Science Park 904, 1098 XH Amsterdam, The Netherlands}

\emailAdd{mary.lizunova@gmail.com}
\emailAdd{vanwezel@uva.nl}

\abstract{
The $\varphi^4$-theory is ubiquitous as a low-energy effective description of processes in all fields of physics ranging from cosmology and particle physics to biophysics and condensed matter theory. The topological defects, or kinks, in this theory describe stable, particle-like excitations. In practice, these excitations will necessarily encounter impurities or imperfections in the background potential as they propagate. Here, we describe the interaction between kinks and various types of realistic impurity models. We find that realistic impurities behave qualitatively like the well-studied, idealized delta function impurities, but that significant quantitative differences appear in both the characteristics of localized impurity modes, and in the collision dynamics. We also identify a particular regime of kink-impurity interactions, in which kinks loose all of their kinetic energy upon colliding with an impurity.
}

\keywords{Field Theories in Lower Dimensions, Solitons Monopoles and Instantons, Effective field theories}

\makeatletter
\def\@fpheader{\relax}
\makeatother
\begin{document} 

\maketitle

\flushbottom

\section{\label{sec:level1} Introduction}
From its introduction in the context of the Ginzburg-Landau theory for second order phase transitions~\cite{gltheory}, $\varphi^4$-theory has found applications in the low-energy description of many physical processes, including domain wall motion~\cite{wada}, molecular dynamics~\cite{stoll}, and chemical equilibrium~\cite{aubry}. Likewise, the solitary wave solutions of $\varphi^4$-theory, known as kinks, feature in the phenomenological theory of domain walls in  molecules, solids, and cosmology~\cite{bishop, bishop2,mele, friedland}, as well as in toy models for nuclear physics~\cite{campbell1, wick, boguta}. Recently, using a gauge field theory approach, it was found that the kinks of $\varphi^4$-theory may even be useful in the description of biologically relevant molecules, where they can describe for example the native structure of bending angles in complex proteins like Myoglobin~\cite{protein1,protein2,protein3}. Collisions among kinks in $\varphi^4$-theory are particularly interesting, because unlike for example solitons in the integrable sine-Gordon model~\cite{aek}, kinks and antikinks cannot just pass through each other. Instead, they interact and undergo dynamic processes including scattering, formation of bound states, and even resonances~\cite{ablowitz, anninos, goodman2, lizunova02,lizunova03}.  

Besides kink-antikink collisions, interactions between kinks and impurities are a central ingredient in the modelling of any realistic system. In chemical and condensed matter settings, actual impurities in the atomic lattice are an unavoidable fact of life, while in particle physics and cosmology variations in the potential or background metric act as impurities~\cite{kivshar,konotop,javidan,javidan01,javidan02,javidan03,askari}. Idealized impurities, with a Dirac delta spatial profile, have been shown to be capable of scattering or capturing kinks, as well as harbouring a localized impurity mode of their own~\cite{kivshar}. Here, we extend these results by including interactions between kinks and more realistic impurity models, based on Gaussian or Lorentzian spatial profiles. We find qualitative agreement with the physics of idealized defects, but significant quantitative effects of the impurity profile on the shape of the localized impurity mode, the scattering dynamics, and the possible long-time fate of kinks interacting with strong impurities.

\section{\label{sec:level2} Scalar fields in (1+1) dimensions}
We consider a classical, real scalar field $\varphi=\varphi(t,x)$ in $(1+1)$-dimensional space-time, described by the Lagrangian density~\cite{bazeia01,radjaraman}:
\begin{equation}\label{eq:largang}
	\mathscr{L}=\frac{1}{2} \left( \frac{\partial\varphi}{\partial t} \right)^2-\frac{1}{2} \left( \frac{\partial\varphi}{\partial x} \right) ^2-\frac{1}{4}(1-\varphi^2)^2.
\end{equation}
The final term represents the self-interaction potential of the field $\varphi$. It has two minima, $\varphi_v^{(1)}$ and $\varphi_v^{(2)}$, defining the vacuum manifold of the theory. From the Euler-Lagrange equation, the field $\varphi$ is found to obey the equation of motion:
\begin{equation}\label{eq:eom_phi4}
\varphi_{tt}-\varphi_{xx}+\varphi^3-\varphi=0.
\end{equation}

The trivial solutions of this equation are $\varphi=0$, corresponding to an unstable local maximum of the potential, and $\varphi=\varphi_v^{(1,2)}=\pm 1$, which are the stable vacuum solutions. Non-trivial solutions may be found by imposing that the field approaches distinct vacua at opposing spatial infinities. The minimal-energy static solution for any given set of spatial boundary conditions is called a BPS saturated configuration~\cite{bps1,bps2,bazeia01}. For the $\varphi^4$-theory defined by eq.~\eqref{eq:largang}, a non-trivial and non-dissipative static BPS-solution connecting $\varphi_v^{(1)}$ and $\varphi_v^{(2)}$ is given by:
\begin{equation}\label{eq:kink}
\varphi(x)=\pm\tanh\left(\frac{x-a}{\sqrt{2}}\right).
\end{equation}
This solution is known as a kink $\varphi_K$ (or antikink $\varphi_{\overline{K}}$) for the plus (minus) sign. Each has a center at $x=a$ and characteristic length $l_K\simeq\sqrt{2}$. Owing to the Lorentz invariance of eq.~\eqref{eq:eom_phi4} the static (anti)kink solution transforms into a dynamical solution under a Lorentz boost:
\begin{equation}\label{eq:kink_boost}
\varphi(t,x)=\pm\displaystyle\tanh\left(\frac{x-a+vt}{\sqrt{2(1-v^2)}}\right),
\end{equation}
where $0<v<1$ is a velocity of the moving (anti)kink in units of the speed of light.

\begin{figure*}[tb]
\begin{center}
\includegraphics[width=\linewidth]{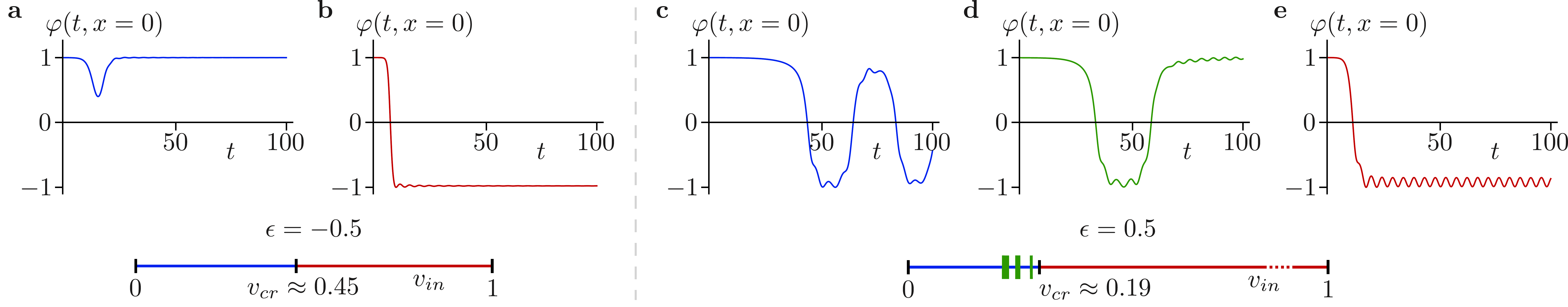}
\end{center}
\caption{The value of the field $\varphi(t,x)$ at $x=0$ as a function of time, for a repulsive impurity with $\epsilon=-0.5$ and $\sigma\simeq 0.016$ (left of the dashed line) and for an attractive impurity with $\epsilon=+0.5$ and $\sigma\simeq 0.016$ (right of the dashed line). {\bf a)} For low initial velocity, $v_{in}=0.4$, the kink is reflected by the impurity. {\bf b)} For high initial velocity, $v_{in}=0.8$, the kink traverses the impurity, at the cost of loosing some kinetic energy. {\bf c)} For $v_{in}=0.1$, the kink is captured by the attractive impurity. {\bf d)} For specific intermediate values such as $v_{in}=0.137$, a resonance causes the kink to oscillate around the impurity before being released again. A few of the resonance windows are sketched along the bar of $v_{in}$ values at the bottom. {\bf e)} For high initial velocity, $v_{in}=0.5$, the kink traverses the impurity, leaving behind an excited impurity mode, and loosing some kinetic energy.  
}
\label{fig:repuls_imp}
\end{figure*}

\section{\label{sec:level3} Impurity}
We introduce a single inhomogeneity $\gamma(x)$ into the potential of  eq.~\eqref{eq:largang} by writing~\cite{kivshar,konotop,javidan03}:
\begin{equation}\label{eq:u_impurity}
\frac{1}{4}(1-\varphi^2)^2 \longrightarrow \frac{1}{4}(1-\varphi^2)^2(1-\epsilon \gamma(x-x_0)).
\end{equation}
Here, $|\epsilon|<1$ and $x_0$ represent the strength and the center of a weak impurity. For $\epsilon=0$ the clean model is recovered, while $\epsilon<0$ and $\epsilon>0$ correspond to a repulsive barrier and an attractive potential well respectively. Including the impurity, the equation of motion becomes:
\begin{equation}\label{eq:eom_impurity}
\varphi_{tt}-\varphi_{xx}+(\varphi^3-\varphi)(1-\epsilon\gamma (x-x_0))=0.
\end{equation}

We numerically solve eq.~\eqref{eq:eom_impurity} using a finite differences method (see appendix for details). The maximal detected fluctuations in energy during time evolution of an initial field configuration were less than $0.74\%$, indicating the stability and accuracy of the numerical routine.

We consider the initial condition of a single kink $\varphi=\varphi_K$, centered at $t=0$ at the position $a=6$, and moving with some initial velocity $v=v_{in}$ towards an impurity located at $x_0=0$. We explore the influence of the impurity profile $\gamma(x)$ on the kink-impurity scattering by considering three distinct types of impurities. The first is the idealized Dirac delta function originally proposed in~\cite{kivshar}:
\begin{equation}\label{eq:imp_type1}
\gamma (x-x_0)\rightarrow\delta (x-x_0).
\end{equation}
We then generalise to a more realistic Gaussian form for the impurity:
\begin{equation}\label{eq:imp_type2}
\gamma (x-x_0)\rightarrow \frac{1}{\sigma\sqrt{2\pi}}\text{exp}\left[-\left(\frac{x-x_0}{\sigma\sqrt{2}}\right)^2\right].
\end{equation}
Here, choosing $0<\sigma<1$ corresponds to the impurity width $\sigma$ being less than the width $l_K$ of the kink. We compare the results of the Gaussian impurity model with a final realistic Ansatz, given by a Lorentzian profile:
\begin{equation}\label{eq:imp_type3}
\gamma (x-x_0)\rightarrow\frac{1}{(x-x_0)^2+\alpha^2}.
\end{equation}

\section{\label{sec:level4} Kink-impurity interactions}

\subsection{\label{sec:level4.1} Dirac delta limit of the Gaussian profile}
To establish a connection to the known result for the idealised Dirac delta impurity~\cite{kivshar}, we first consider a very narrow form of the Gaussian profile. We choose the height of the Gaussian at its centre to coincide with the numerical value for the delta function height used in ref.~\cite{kivshar}. This results in a width $\sigma\simeq 0.016$ that is smaller than the lattice spacing in our numerical routine, in accordance with the Dirac delta limit.

For a repulsive impurity with $\epsilon= -0.5$, we reproduce the two types of processes known to occur as the value of $v_{in}$ is varied~\cite{kivshar}. For low initial velocities, such as $v_{in}=0.4$ (shown in figure~\ref{fig:repuls_imp}a), the kink is reflected by the impurity. For velocities above some critical value $v_{cr}$, the kink is transmitted, but its kinetic energy (velocity) is reduced in the process and emitted in the form of low-amplitude ripples. This is shown in figure~\ref{fig:repuls_imp}b for the initial velocity $v_{in}=0.8$.

In the case of an attractive impurity, with $\epsilon=+0.5$, three different types of known dynamics are reproduced~\cite{kivshar}. At low initial velocities, such as $v_{in}=0.1$ (shown in figure~\ref{fig:repuls_imp}c), the kink is captured by the impurity. That is, it ends up in a final state in which the kink centre oscillates around the central position of the impurity. For high initial velocities on the other hand, like $v_{in}=0.5$ (shown in figure~\ref{fig:repuls_imp}e), the kink is transmitted through the impurity entirely, loosing some kinetic energy in the process, and leaving behind a localized excitation centered at the impurity location. The localized impurity mode consists of small amplitude periodic oscillations of the field at $x=x_0$ around its vacuum value $\varphi_v^{(1)}$. The value of the initial velocity above which kinks are always transmitted through the impurity is called the critical velocity $v_{cr}$. For certain particular values of the initial velocity $v_{in}<v_{cr}$, such as $v_{in}=0.137$ (shown in figure~\ref{fig:repuls_imp}d), the kink is neither captured nor transmitted. Instead, it oscillates around the impurity center a finite number of times before leaving the impurity in the direction or opposite one it originally came from. This type of behavior is known as resonance, and can be understood in terms of the exchange of kinetic energy with internal excitations of the kink~\cite{kivshar,Campbell}.

For both attractive and repulsive impurity strengths, our results are in perfect agreement with those reported in ref.~\cite{kivshar}, confirming that the idealized Dirac delta impurity gives a faithful approximation of the more realistic Gaussian profile in the limit of very low width. Notice that the precise value of $v_{cr}$ we find is slightly higher than that reported before. This is a direct consequence of our numerical routine being stable to higher final times than previously achievable. We observe that at these later times, kinks that seemingly escaped the impurity, still return and are captured. Since we have no way of establishing the dynamics at even later times, we refrain from reporting any more precise value for $v_{cr}$, and instead focus below on its qualitative behavior as the strength and width of the impurity profile are varied.

\begin{figure}[tb]
\begin{center}
\includegraphics[width=\linewidth]{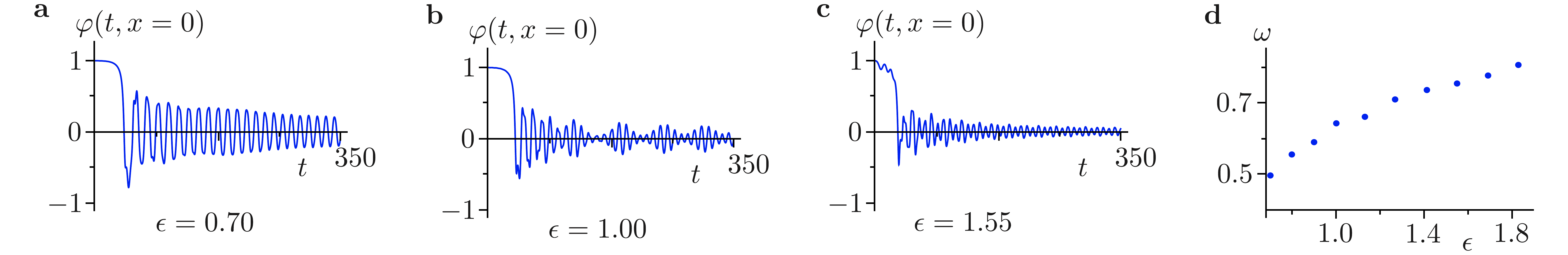}
\end{center}
\caption{{\bf a-c)} The value of the field $\varphi(t,x)$ at $x=0$ as a function of time, for an attractive impurity with width $\sigma\simeq 0.016$, and fixed initial velocity of the kink $v_{in}=0.1$. {\bf a)} For impurity strength $\epsilon=0.7$; {\bf b)} with $\epsilon=1.0$; and {\bf c)} for a very strongly attractive impurity with $\epsilon=1.55$. In all cases the kink is captured by the impurity and oscillates around the impurity centre with an amplitude lower than the kink width. {\bf d)} The dominant frequency $\omega$ of oscillations of the captured kink as a function of impurity strength.
}
\label{fig:sec4.3_5a}
\end{figure}

Finally, we consider the limit of a very strong attractive impurity, with $\epsilon>1$. The resulting dynamics is presented in figure~\ref{fig:sec4.3_5a}a-c. As the impurity strength increases, the captured kink oscillates around the impurity with ever lower amplitude, eventually getting stuck at the impurity position entirely. This regime, in which the oscillation amplitude of a captured kink is smaller than its width, so that the field value at the impurity location never returns to either of its vacuum values $\varphi_v^{(1,2)}$, may be called a super-capture. figure~\ref{fig:sec4.3_5a}d shows the dependence of the dominant oscillation frequency on impurity strength.

\subsection{\label{sec:level4.2} Impurity strength and width}
For an ideal Dirac delta impurity, the value of $v_{cr}$ is determined entirely by the impurity strength $\epsilon$. For vanishing strength, the critical velocity tends to zero, while strong impurities cost a lot of kinetic energy to traverse, yielding a large $v_{cr}$. In the case of a Gaussian impurity profile, the width $\sigma$ of the impurity as well as its strength $\epsilon$ may be expected to play a role in determining the critical velocity, i.e. $v_{cr}=v_{cr}(\epsilon,\sigma)$.

\begin{table}[b]
 \begin{tabular}{c c c c} 
 $\epsilon$ & $\sigma$ & $v_{in}$ & fate of kink \\
 \hline
 0.3 & 0.5 & 0.02 & transmitted \\
 0.3 & 0.3 & 0.02 & captured \\
 0.3 & 0.3 & 0.05 & transmitted \\
 0.3 & 0.1 & 0.05 & captured \\
 ~\\
 -0.6 & 0.5 & 0.47 & transmitted \\
 -0.6 & 0.3 & 0.47 & reflected \\
 -0.6 & 0.3 & 0.5 & transmitted \\
 -0.6 & 0.1 & 0.5 & reflected
\end{tabular}
\hfill
 \begin{tabular}{c c c c} 
 $\epsilon$ & $\sigma$ & $v_{in}$ & fate of kink \\
 \hline
 0.4 & 0.3 & 0.1 & transmitted \\
 0.6 & 0.3 & 0.1 & captured \\
 0.6 & 0.3 & 0.28 & transmitted \\
 0.8 & 0.3 & 0.28 & captured \\
 ~\\
 -0.1 & 0.3 & 0.3 & transmitted \\
 -0.4 & 0.3 & 0.3 & reflected \\
 -0.4 & 0.3 & 0.5 & transmitted \\
 -0.8 & 0.3 & 0.5 & reflected 
\end{tabular}
\caption{{\bf Left)} The effect of varying impurity width on the kink-impurity interaction, for fixed value of the impurity strength. Larger values of $\sigma$ are seen to yield lower values of $v_{cr}$. {\bf Right)} The effect of varying impurity strengt, for fixed value of the impurity width. Larger values of $|\epsilon|$ are seen to yield higher values of $v_{cr}$.
}
\label{table1}
\end{table}

The qualitative effects of the strength and impurity may be understood from the results in table~\ref{table1}. The critical velocity increases in value upon either decreasing the width $\sigma$ for fixed values of the strength $|\epsilon|$, or upon increasing $|\epsilon|$ for fixed values of $\sigma$. These trends can be understood by comparing eq.~\eqref{eq:eom_impurity} and eq.~\eqref{eq:imp_type2}, which show that the height of the impurity potential at $x_0$ is proportional to the ratio $\epsilon/\sigma$.

To further confirm that the interaction of kinks with a wide impurity differs only quantitatively from that with a narrow impurity, we confirm that resonances still occur for realistic impurity widths. For the values $\epsilon=0.5$ and $\sigma\simeq 0.016$ there is a resonance around $v_{in}=0.137$. Upon increasing the impurity width to $\sigma=0.1$, we instead find a resonance at $v_{in}=0.15$. At $\epsilon=0.6$ and $\sigma=0.3$ we observe a resonance at the same value of $v_{in}$. The qualitative behavior of having resonance windows below $v_{cr}$ for attractive impurities thus survives also for more realistic impurity profiles of non-zero width.

\subsection{\label{sec:level4.3} The impurity mode}
\begin{figure}[tb]
\begin{center}
\includegraphics[width=\linewidth]{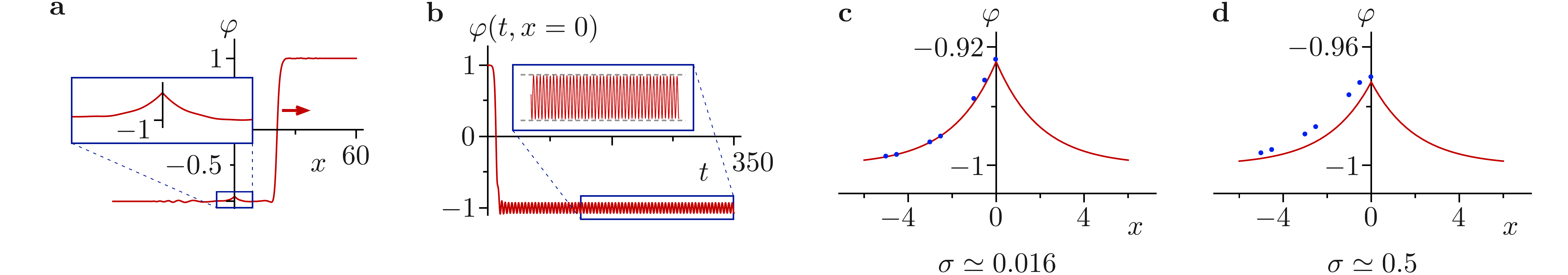}
\end{center}
\caption{{\bf a)} The spatial profile of the impurity mode, excited by a kink traversing the impurity site $x=x_0=0$ and travelling to the right. {\bf b)} The value of the field $\varphi(t,x)$ at $x=x_0=0$ as a function of time, for an attractive impurity with width $\sigma= 0.5$ and strength $\epsilon=0.5$. An impurity mode is excited by a kink with initial velocity $v_{in}=0.5$ passing $x=x_0$ at approximately $t\approx 20$. The horizontal dashed lines in the inset are guides to the eye highlighting the quasi-long-lived character of the impurity mode. {\bf c)} Comparison of the numerically obtained values for the field $\varphi(t,x)$ after a kink traversed the impurity site (blue dots), with the  analytical prediction of eq.~\eqref{eq:impurity_mode} (red line). The impurity potential used is a narrow Gaussian with strength $\epsilon=0.5$ and width $\sigma\simeq0.016$, while the incoming kink had initial velocity $v_{in}=0.5$. {\bf d)} The same comparison using a wide Gaussian impurity potential with strength $\epsilon=0.5$ and width $\sigma\simeq0.5$.
}
\label{fig:sec4.3_3}
\end{figure}

When a kink traverses an attractive impurity it leaves behind an oscillating mode, localized at the site of the impurity (see figure~\ref{fig:sec4.3_3}a). This mode oscillates at $x=x_0$ with near-constant amplitude and frequency, as shown in figure~\ref{fig:sec4.3_3}b. It is a quasi-long-lived mode, analogous to the wobbling kink~\cite{lizunova01}, and bion solutions found in kink-antikink collisions~\cite{aek}.

For the case of the idealised Dirac delta impurity, the approximate shape of the impurity mode profile can be found analytically~\cite{kivshar}. Since the mode is localized at the impurity site and excited by the passing of a kink, we can write it as a small deviation $\delta\varphi$ from the vacuum solution. Substituting $\varphi(t,x)=\varphi_v^{(1)}+\delta\varphi(t,x)$ in eq.~\eqref{eq:eom_impurity} and eq.~\eqref{eq:imp_type1} and keeping only terms up to linear order in $\delta\varphi$ yields the equation of motion for the impurity mode:
\begin{equation}
    \delta\varphi_{tt}-\delta\varphi_{xx}+2\left(1-\epsilon\delta(x-x_0)\right)\delta\varphi = 0.
\end{equation}
Using the Ansatz that the impurity mode oscillates with a fixed spatial profile and constant frequency, $\delta\varphi(t,x)\propto\Re~\chi(x)\text{exp}(-i\Omega t)$, it is found to obey~\cite{kivshar}:
\begin{equation}\label{eq:impurity_mode}
    \delta\varphi(t,x)\propto\Re~\text{exp}\left(-\epsilon|x|\right)\text{exp}(-i\Omega t),\quad  \Omega^2=2-\epsilon^2.
\end{equation}

The numerically obtained field configuration in figure~\ref{fig:sec4.3_3}c indicates that for a very narrow impurity potential, $\epsilon=0.5$ and $\sigma\simeq0.016$, an incoming kink with $v_{in}=0.5$ excites an impurity mode whose shape closely matches the prediction of eq.~\eqref{eq:impurity_mode}. The numerical results in figure~\ref{fig:sec4.3_3}d on the the other hand, show that for wider impurity potential, $\epsilon=0.5$ and $\sigma\simeq0.5$, the analytic solution no longer gives an accurate prediction for the impurity mode profile. The spatially smooth impurity potential in this case does not allow for any discontinuities in the field configuration or its derivatives, forcing the impurity mode profile to remain smooth around the impurity location.

\begin{figure}[tb]
\begin{center}
\includegraphics[width=\linewidth]{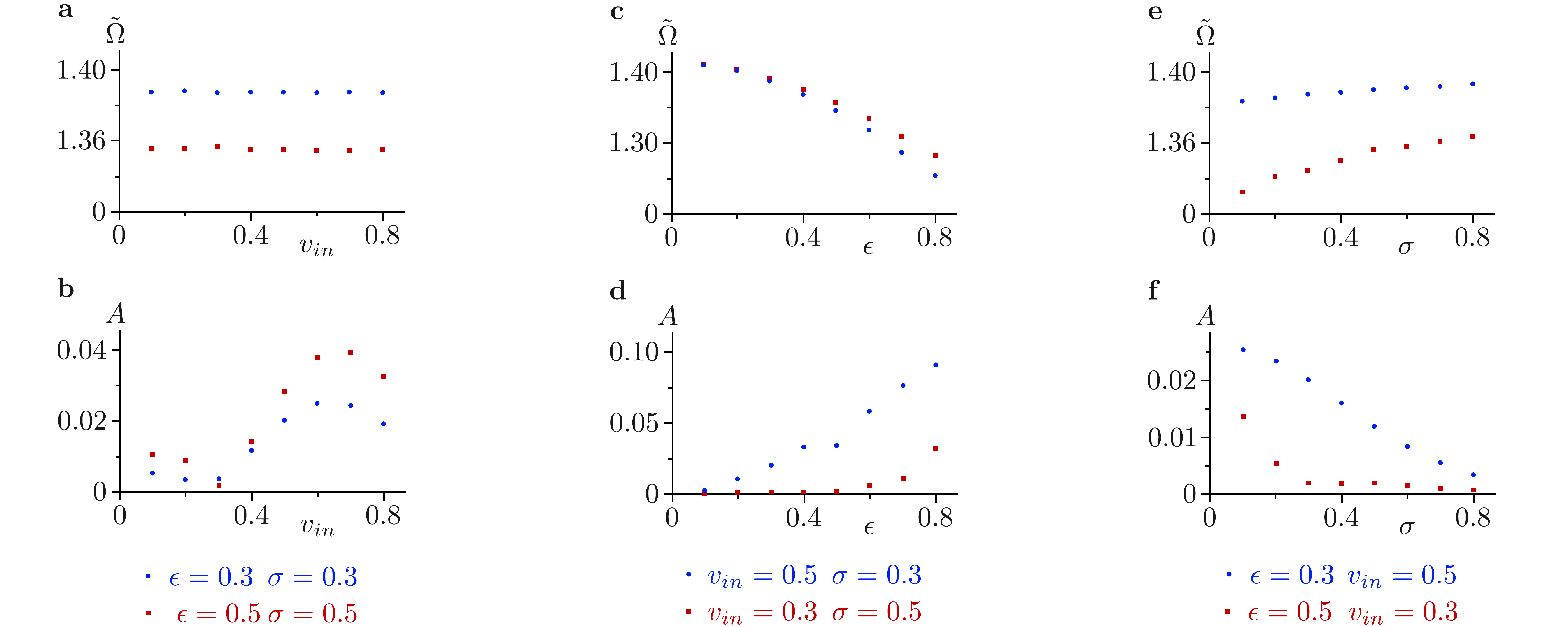}
\end{center}
\caption{{\bf a)} The impurity mode frequency $\tilde\Omega$ and {\bf b)} its amplitude $A$, as a function of the initial kink velocity $v_{in}$, for fixed values of impurity strength and width. {\bf c)} The frequency and {\bf d)} amplitude as a function of impurity strength $\epsilon$, keeping the initial velocity and impurity width constant. {\bf e)} Frequency and {\bf f)} amplitude as a function of the the impurity width $\sigma$, for fixed impurity strength and initial velocity.
}
\label{fig:sec4.3_2a}
\end{figure}

Besides the spatial profile of the impurity mode, its amplitude $A$ and frequency $\tilde \Omega$ may also vary with impurity width, and deviate from the analytic prediction of eq.~\eqref{eq:impurity_mode}. For narrow impurity potential ($\sigma\simeq0.016$) we find $\langle \tilde\Omega \rangle \simeq 1.382$ for $\epsilon=0.3$ and $\langle \tilde\Omega \rangle \simeq 1.323$ for $\epsilon=0.5$, both in excellent agreement with the analytic prediction of eq.~\eqref{eq:impurity_mode}. For wider impurities, the approximations underlying the analytic solution break down, and we numerically find the behavior displayed in figure~\ref{fig:sec4.3_2a}. 

The frequencies in figure~\ref{fig:sec4.3_2a}b do not vary much with the initial velocities. Taking the average over all points shown yields a value of $\langle\tilde\Omega\rangle\simeq 1.387$ for $\epsilon=\sigma=0.3$, and $\langle\tilde\Omega\rangle\simeq 1.355$ for $\epsilon=\sigma=0.5$. The decrease of the frequency with increasing impurity strength indicated in figure~\ref{fig:sec4.3_2a}d is in line with the analytic prediction of eq.~\eqref{eq:impurity_mode} for ideal Dirac delta impurities, but is now seen to also depends on the width $\sigma$ of the more realistic Gaussian profile. The frequency increases with increasing width, as shown in figure~\ref{fig:sec4.3_2a}f, in agreement with the fact that increased width in eq.~\eqref{eq:imp_type2} implies a lower value of the impurity potential at $x=x_0$. 

\subsection{\label{sec:level4.4} Lorentzian profile}
To see whether the qualitative effects of a wide impurity profile, rather than an idealised Dirac delta form, are generic to more realistic impurity models, we next compare the results of the Gaussian case to interactions between a kink and a Lorentzian impurity. We repeat the analysis of section~\ref{sec:level4.1} using the impurity profile of eq.~\eqref{eq:imp_type3}, with $\epsilon=\pm 0.5$. To ensure that the peak height at $x_0=0$ matches that of the Gaussian considered before, we take $\alpha=0.2$. 

For the repulsive Lorentzian impurity, we find that the kink is reflected for all initial velocities up to $v_{in}=0.9$. However, at $v_{in}=0.99$ the kink traverses the impurity, showing that a critical velocity exists in the range $0.9<v_{in}<0.99$. Reversing the sign of $\epsilon$ and considering an attractive Lorentzian impurity, we observe that for initial velocities below $v_{in}=0.8$, the kink is always captured by the impurity.

Although the capturing and reflecting of kinks is qualitatively similar to the behavior in the presence of a Gaussian impurity, the values of the critical velocities are much higher for a Lorentzian profile than for a Gaussian with the same maximum value. We therefore also compare Lorentzian and Gaussian profiles with the same integrated strength, by using the scaled function:
\begin{equation}
    \frac{\alpha/\pi}{(x-x_0)^2+\alpha^2}.
\end{equation}
We again take $\epsilon=\pm 0.5$ and $\alpha=0.2$.

In this case, the kink is reflected by a repulsive impurity for initial velocities up to $v_{in}=0.4$, and traverses the impurity site for velocities $v_{in}=0.5$ and above. Likewise, the attractive impurity captures the kink at $v_{in}=0.01$, while it is transmitted for initial velocities above $v_{in}=0.1$.

These results indicate a qualitative agreement between the dynamics of kink-impurity interactions for Gaussian and Lorentzian impurity profiles. We leave more detailed analyses for future study, looking for example for resonances around the Lorentzian impurity, or investigating the precise influence of the parameter $\alpha$ on the critical velocity.

\section{\label{sec:conclusions} Conclusion}
In conclusion, we have shown that the behavior of kink-impurity interactions in 
$(1+1)-$dimensional $\varphi^4$-theory is qualitatively the same for idealised Dirac delta impurity profiles and more realistic Gaussian or Lorentzian shapes. In all cases, repulsive impurities can either reflect or transmit an incoming kink, depending on its initial velocity. Attractive impurities on the other hand either capture the kink, release after a few oscillations in a resonant process, or transmit it immediately. In the latter case, an impurity mode is excited at the impurity location by the passing kink. We also observe that a kink impinging on particularly strong impurities may lead to an extreme form of capture, in which the kink looses all of its kinetic energy and remains localized at the impurity site.

The Gaussian profile reduces to precisely a Dirac delta impurity in the limit of vanishing width, and all observed properties of the kink-impurity interaction agree with the known results for ideal impurities in that limit. For wider impurity profiles, the results differ quantitatively from the ideal case. 
The amplitude and frequency of the impurity mode obtain a dependence on the impurity width, the values of critical velocities are affected, and the kink velocity at which resonances appear is altered. Similar quantitative effects are observed for interactions between a kink and a Lorentzian impurity. 

\acknowledgments 
The authors are very grateful for discussions with Dario Bercioux, Alexander Kudryavtsev, and Cristiane Morais Smith. This work was done within the Delta Institute for Theoretical Physics (DITP) consortium, a program of the Netherlands Organization for Scientific Research (NWO) that is funded by the Dutch Ministry of Education, Culture and Science (OCW).

\appendix
\section{Numerical methods}
\subsection{Integration}
To integrate the equation of motion in eq.~\eqref{eq:eom_impurity}, we use finite differences on a discretized lattice in both space and time:
\begin{align}\label{eq:eom_scheme}
\varphi_j^{k+1} = 2\varphi_j^k-\varphi_j^{k-1}+\frac{\tau^2}{h^2}(\varphi_{j+1}^{k}-2\varphi_j^k+\varphi_{j-1}^{k}) -\tau^2 \left(\varphi^3-\varphi\right)_j^k(1-\epsilon\gamma(x_j-x_0)).
\end{align}
Here, $\varphi_j^k$ equals $\varphi(x,t)$ at the discrete positions $x_j=-L+jh$ and times $t_k=k\tau$, with $\tau$ and $h$ the sizes of the discrete space and time steps. We demand $\tau<h$ to ensure stability of the numerical integration.

All calculations reported in this article use $0<t<t_f$ with $t_f=350$, and $-L<x<L$ with $L=730$. To avoid any effect of the spatial boundaries on the reported results, we present only results within the interval $-L'<x<L'$, where $L'=L - t_f h/\tau$.

As a measure of the accuracy of the numerical integration, we check whether energy is conserved in time, taking into account the flow of energy through the borders $x=\pm L'$ at each time step $t_k$:
\begin{equation}\label{eq:law_energycon}
E[\varphi(t=t_k)]-\int\limits_{0}^{t_k} \frac{\partial\varphi}{\partial t}\frac{\partial\varphi}{\partial x} \biggl|_{-L'}^{L'} dt=E[\varphi(t=0)].
\end{equation}
In this expression, the instantaneous energy is found from the spatial integral of the Lagrangian, $E[\varphi]=\int \mathscr{L}[\varphi] \, dx$. Appropriate values for the steps $\tau$ and $h$ are determined empirically by demanding that energy is conserved, and all results are independent of the chosen step values. Here, we use $\tau=0.01$ and $h=0.02$, and the maximal detected deviation in energy is less $0.74\%$.

\subsection{Frequency and amplitude}
To extract the frequency and amplitude of the impurity mode oscillations from the numerically obtained field profile, we can either use a discrete Fourier transform of $\varphi(x_0,t)$, or directly average the distances between its observed minima and maxima. In either approach, we only consider the field values for $t>170$, to ensure that the kink has entirely passed by the impurity centre. The accuracy of both methods is limited by the discrete sampling of the continuous field $\varphi(x_0,t)$, and the values obtained for $A$ and $\tilde \Omega$ with the two methods do not differ significantly.

\bibliographystyle{JHEP}
\bibliography{bibliography}

\end{document}